\documentclass[
superscriptaddress,
 amsmath,amssymb,
 aps,
prb,two column
]{revtex4-1}

\usepackage{graphicx}
\usepackage{dcolumn}
\usepackage{bm}

\usepackage{graphicx,epsf,bm,bbm,color,amsmath,ulem,amssymb,float, subfigure}

\usepackage{color}
\usepackage{ textcomp }
\usepackage{ dsfont }
\usepackage{indentfirst}
\newcommand{\remove}[1]{}

\newcommand{\bi}{\begin{itemize}}
\newcommand{\ei}{\end{itemize}}
\newcommand{\be}{\begin{enumerate}}
\newcommand{\ee}{\end{enumerate}}

\newenvironment{dfn}{{\vspace*{1ex} \noindent \bf Definition }}{\vspace*{1ex}}

	\newcommand{\beq}{\begin{eqnarray}}
	\newcommand{\eeq}{\end{eqnarray}}

\begin{document}
\title{Classification and Distinction of Possible Insulating Phases in Twisted Bilayer Graphene by Impurity Effects}

\author{Zhi-Qiang Gao}
\affiliation{International Center for Quantum Materials, School of Physics, Peking University, Beijing 100871, China
}

\author{Fa Wang}
\affiliation{International Center for Quantum Materials, School of Physics, Peking University, Beijing 100871, China
}
\affiliation{Collaborative Innovation Center of Quantum Matter, Beijing 100871, China
}

\date{\today}

\begin{abstract}
In this work the effects of impurity in various insulating phases of the twisted bilayer graphene (TBG) are studied. The well-accepted continuum model\cite{b} is employed and the local density of states (DOS) is calculated. It is found that insulating phases breaking different symmetries proposed in previous theories\cite{pa,op1,Lee_2019} are distinguishable via the number and properties of in-gap bound state peaks induced by impurities in local DOS. Insulating phases breaking the same previously proposed symmetries can be further classified by the remaining anti-unitary symmetries and distinguished by the corresponding remaining Kramers degeneracy of bound states. The in-gap bound state peaks in local DOS and the degeneracy of the bound states can in principle be detected in scanning tunnelling microscopy (STM) experiments, and thus can help to the distinction of various insulating phases.
\end{abstract}


\maketitle

\section{Introduction}

Twisted bilayer graphene has attracted much interest recently. When the twisted angle\cite{b} $\theta \approx 1.08^{\circ}$, flat bands will emerge, which can give rise to rich physics\cite{Cao_2018,Cao_2018a,pa,op1,Lee_2019,Xie:2020aa,H,Ochi:2018aa,Ra:2018aa,Cao:2016aa,S:2012aa,Khalaf:2020aa,Lewandowski:2020aa,Nuckolls:2020aa,Hejazi:2020aa,Christos:2020aa,Zhang:2020ab,Khalaf:2020ab,Cao:2020ab,Ledwith:2020aa,Yang:2019aa,10.21468/SciPostPhys.7.4.048,Serlin:2020aa,Park:2019aa,Wu:2020aa,Khalaf:2019aa,Ramires:2019aa,Wu:2019ab,Zhang:2019aa,Yuan_2019,Cao:2020aa,Zhu:2019aa,Seo:2019aa,Lian:2020aa}. At charge neutral point when the flat bands are half filled, an insulating phase is discovered\cite{Cao_2018}. Various kinds of insulating phases have been proposed\cite{pa,op1,Lee_2019,Khalaf:2020aa,Xie:2020aa,H,Ochi:2018aa,Ra:2018aa,Pi_2018,Cao:2016aa,S:2012aa} to explain this observation. However, the order parameter of the insulating phase is not determined in experiments yet. This work aims to find an indicator to distinguish these different proposed insulating phases in experiments.

Impurities inevitably exists in graphene\cite{Neto:2009aa}. A single impurity, non-magnetic or magnetic, may induce bound states in insulating phases of the TBG. In this work we find that for insulating phases with different order parameters, the degeneracy of the bound states may be different. Therefore, the degeneracy of bound states can serve as the indicator to help to the distinction of the insulating phases in STM experiments. This paper is organized as follows. In Sec. II, the model employed\cite{b,pa,op1,Lee_2019} is briefly reviewed. In Sec. III, the local DOS for various insulating phases proposed in Ref.\cite{pa,op1,Lee_2019} is calculated and we find that insulating phases breaking different symmetries proposed in previous theories are distinguishable via local DOS. In Sec. IV a classification of the insulating phases is given by the remaining Kramers degeneracy of bound states corresponding to remaining anti-unitary symmetries in different physical conditions. The conclusions are drawn in Sec. V. The details of calculation are left in Appendix.

\section{The Model}

To describe the electronic structure of the twisted bilayer graphene (TBG), the model proposed by Ref.\cite{b} is employed and the hybridization of ten $K$ or $K'$ points of Moiré Brillouin zone (MBZ) is considered, as shown in the FIG. 1(a). The region of integration in the calculation is also shown in FIG. 1(a). The band structure is shown in FIG. 1(b), where the flat bands are coloured as red. 
\begin{figure}[htbp]
\centering
\includegraphics[width=8cm]{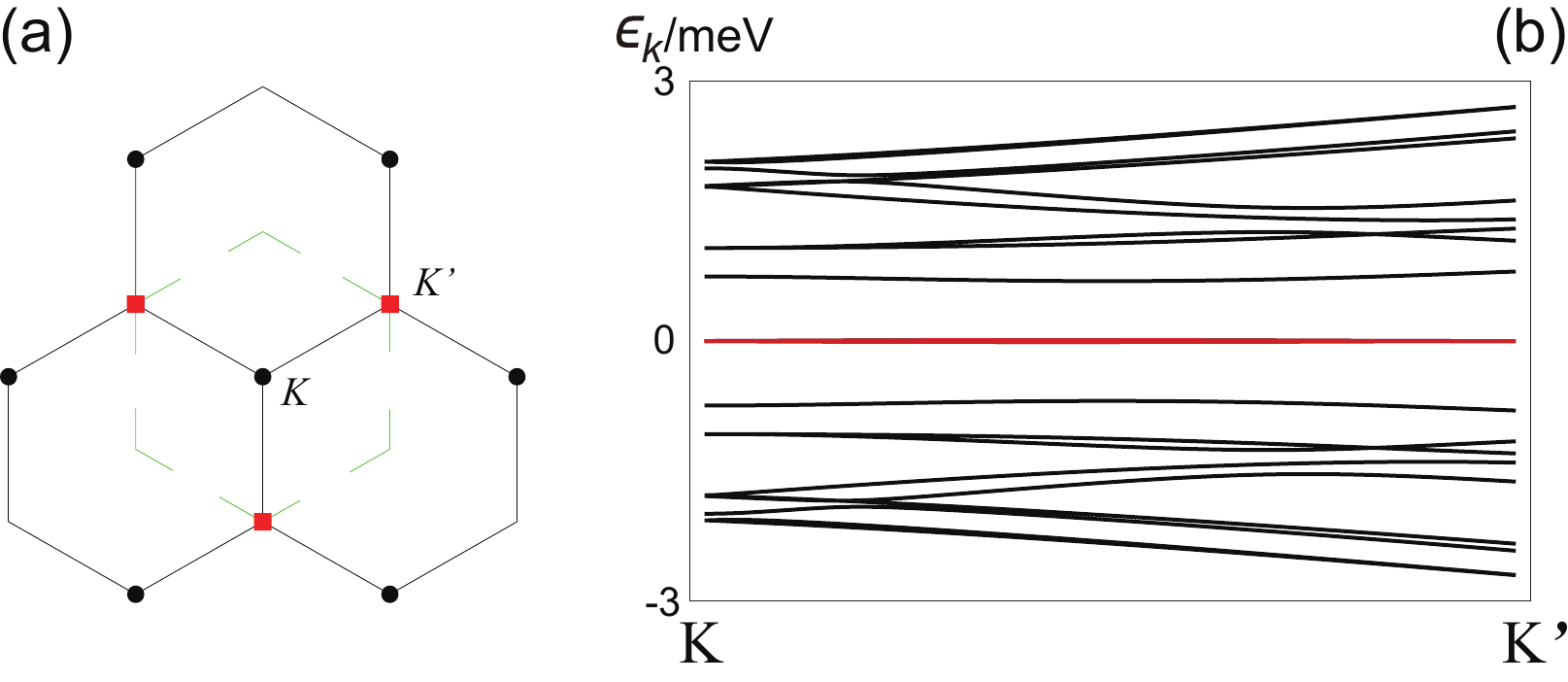}
\caption{(a) Reciprocal lattice. $K$ and $K'$ points included in calculation are marked as black and red dots, respectively. The region of momentum integration is bounded by the green dashed line. (b) The band structure. The two flat band are coloured as red and each has a 4-fold spin-valley degeneracy.}
\end{figure}

The flat band Hamiltonian given by this model\cite{b} reads
\beq
H_0(\vec{k})&=&\sigma^0 \tau^0 \gamma^z \epsilon(\vec{k}),\label{eq:31}
\eeq
where $\sigma$, $\tau$ and $\gamma$ are Pauli matrices in spin, valley and sublattice space, respectively, and $\epsilon(\vec{k})$ is the dispersion of a flat band with $\vec{k}$ measured from a $K$-point in the MBZ. The non-magnetic short-ranged impurity potential in a single layer is $U_{\text{imp}}(\vec{r})=u\delta_{\vec{r},\vec{R}_{0}}$, where $u$ is the strength of the potential and $\vec{R}_{0}$ is the location of the impurity in a single layer. In this work, the results only depend on the short-range-ness and the symmetry in spin-valley space of the impurity potential. Thus this simple $\delta$-function potential is believed to be sufficient for our purpose. The impurity potential projected to the flat bands reads
\beq
H_{\text{imp}}(\vec{k},\vec{k}')&=&u\sigma^0(\tau^0 \text{Re }U(\vec{k},\vec{k}')+i\tau^z \text{Im }U(\vec{k},\vec{k}'))\label{eq:104},
\eeq
where the two by two matrix $U(\vec{k},\vec{k}')$ is the impurity Hamiltonian in sublattice space. The details of this calculation are listed in Appendix A. For an insulating phase, the order parameter $M$ given by a mean-field theory reads
\beq
M=g\sigma^{\mu} \tau^{\nu} \gamma^{\lambda} ,
\eeq
where $\mu,\nu,\lambda=0,x,y,z$, determined by certain symmetry of the insulating phase and $g$ is the amplitude of the order parameter. Several order parameters are proposed by Ref\cite{Lee_2019,op1,pa}, named IVC (Inter-Valley Coherent), SIVCL (Spin-IVC Locked), VH (Valley Hall), QH (Quantum Hall), VSH (Valley Spin Hall), SH (Spin Hall), VP (Valley Polarized), SP (Spin Polarized) and SVL (Spin-Valley Locked). The total Hamiltonian reads
\beq
H(\vec{k},\vec{k}')=H_0(\vec{k})+M+H_{\text{imp}}(\vec{k},\vec{k}').\label{eq:32}
\eeq

In all the calculations the amplitude of order parameters is set to be $g=0.4$ meV to make it much larger than the band width\cite{pa}. For VP, SP and SVL phase\cite{Lee_2019} the strength of the impurity is $u=0.001$ meV since the bound states only emerge when $u$ is small, while for the other phases $u=1$ meV. We have checked that when the values of $g$ and $u$ satisfy that $g$ is much larger than the band width and bound states can emerge in the insulating gap, the results do not qualitatively change.

\section{Local Density of States}

\begin{figure}[htbp]
\centering
\includegraphics[width=9cm]{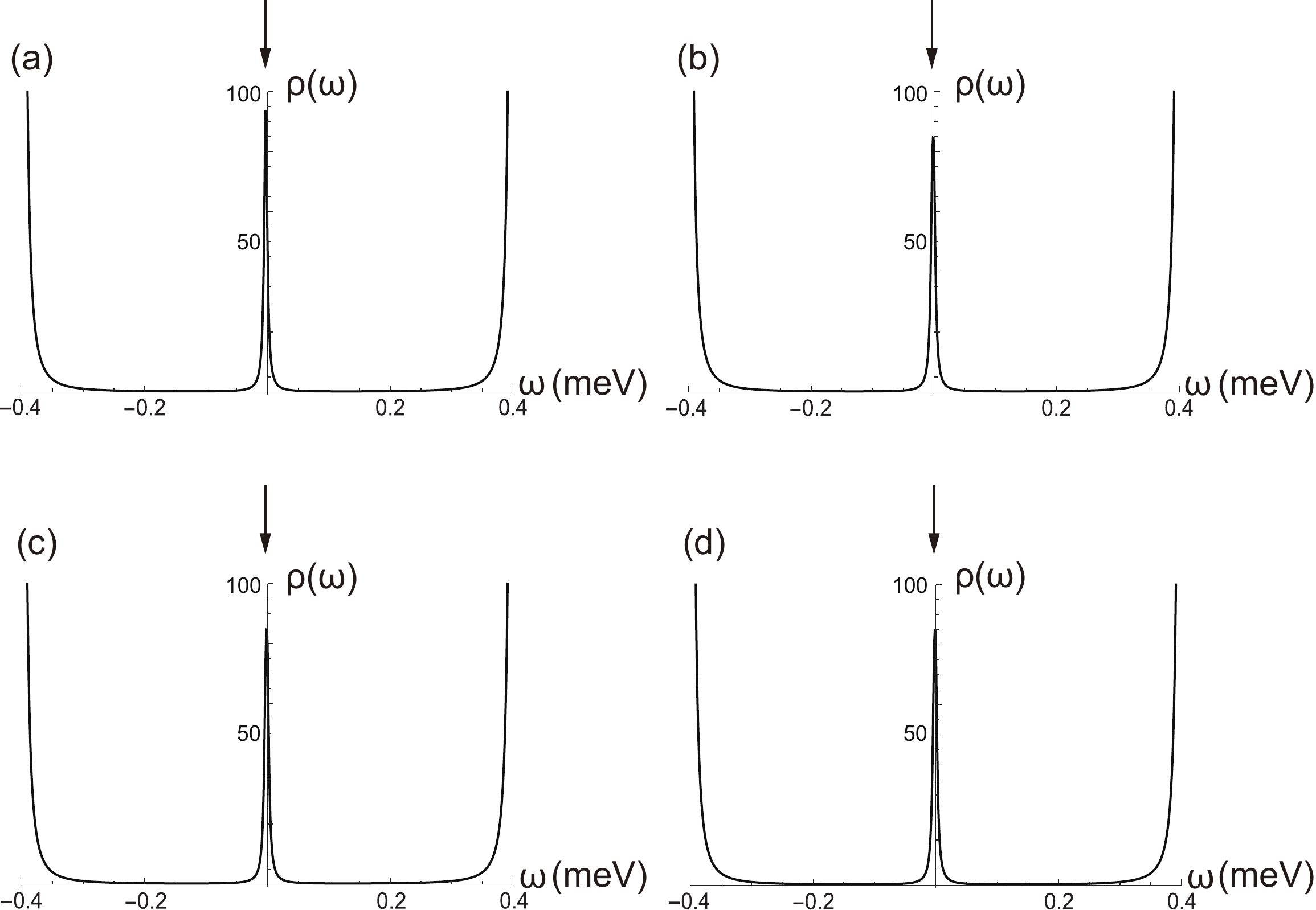}
\caption{Local DOS for phases breaking $C_2\mathcal{T}$ symmetry. (a). VH phase\cite{op1}. (b). QH phase\cite{op1}. (c). VSH phase\cite{op1}. (d). SH phase\cite{op1}. The bound states are pointed by arrows. Only in VH phase the bound states are 4-fold degenerate. In other 3 phases, the bound states are actually 2 pairs of 2-fold degenerate bound states.}
\end{figure}
\begin{figure}[htbp]
\centering
\includegraphics[width=9cm]{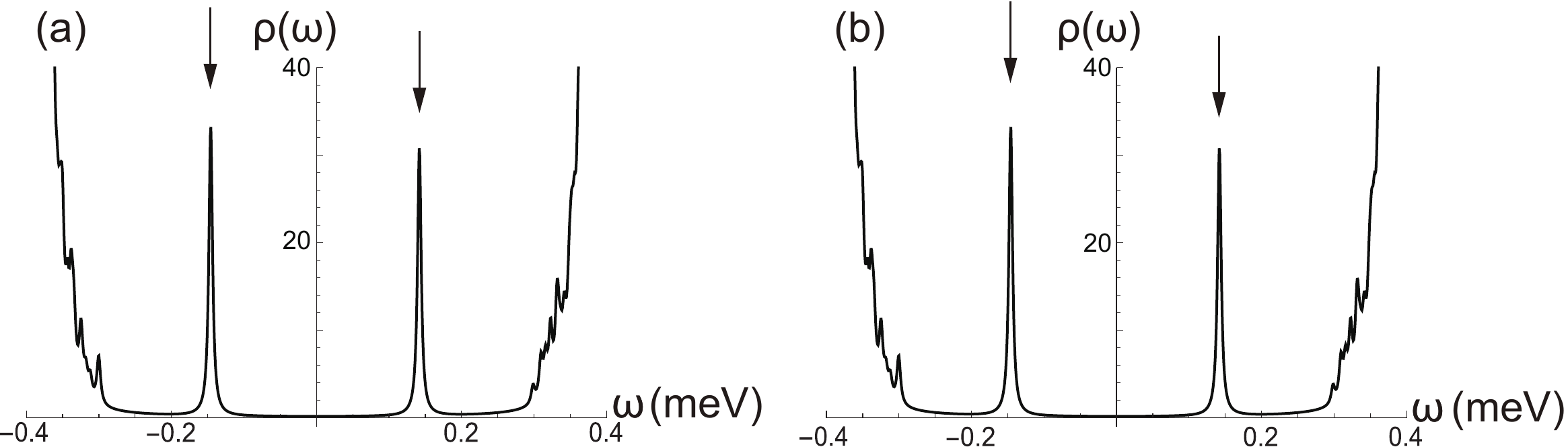}
\caption{Local DOS for phases breaking $U(1)_v$ symmetry. (a). IVC phase\cite{pa}. (b). SIVCL phase\cite{Lee_2019}. All of the bound states indicated by arrows are 2-fold degenerate.}
\end{figure}
\begin{figure*}[htbp]
\centering
\includegraphics[width=13.5cm]{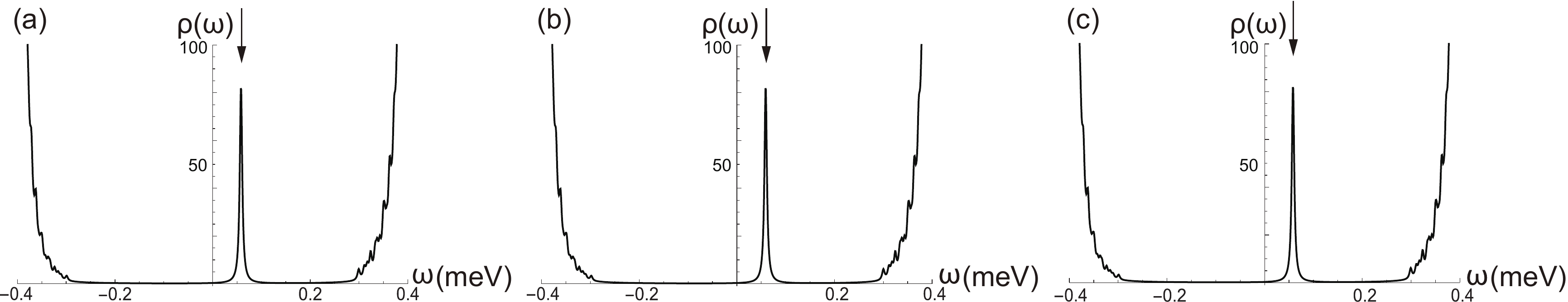}
\caption{Local DOS for phases preserving both $C_2\mathcal{T}$ and $U(1)_v$ symmetries. (a). SP phase\cite{Lee_2019}. (b). VP phase\cite{Lee_2019}. (c). SVL phase\cite{Lee_2019}. All of the bound states indicated by arrows are 2-fold degenerate.}
\end{figure*}

The local DOS of insulating phases breaking $C_2\mathcal{T}$ symmetry\cite{op1}, phases breaking $U(1)_v$ symmetry\cite{pa}, and phases preserving both $C_2\mathcal{T}$ and $U(1)_v$ symmetry are shown in FIG. 2, FIG. 3 and FIG. 4, respectively. These figures show that the local DOS is only sensitive to the symmetry broken by the insulating phase, but not sensitive to the difference between phases breaking the same symmetry. In insulating phases breaking $C_2\mathcal{T}$ symmetry (VH, QH, VSH and SH phase\cite{op1}), a single non-magnetic impurity will induce two bound state peaks. In insulating phases breaking $U(1)_v$ symmetry (IVC and SIVCL phase\cite{pa}), one bound state peak will be induced. In insulating phases preserving both $C_2\mathcal{T}$ and $U(1)_v$ symmetries (VP, SP and SVL phase\cite{Lee_2019}), the only one bound state peak only emerges when the strength of impurity potential is small compared with the amplitude of the order parameters. Therefore, in conclusion, the number and properties of bound state peaks are different for insulating phases breaking different symmetries previously proposed\cite{pa,op1,Lee_2019}. However, insulating phases breaking the same previously proposed symmetry cannot be distinguished by this criterion.

\section{Classification and Distinction of Insulating Phases}

\subsection{Classification by Remaining Anti-Unitary Symmetries}

A salient result in the local DOS of insulating phases is that all the bound states are at least 2-fold degenerate. This degeneracy is protected by the remaining anti-unitary symmetries $\mathcal{S}$ satisfying
\beq
\left[ H,\mathcal{S}\right]=0,\quad \mathcal{S}^2=-1.
\eeq
The classification via remaining anti-unitary symmetries $\mathcal{S}$ corresponding to different insulating phases is listed in the TABLE I. A detailed numerical calculation to explicitly show the degeneracy of bound states is listed in Appendix B. In VH phase, there is a group of 4-fold degenerate bound states. The origin of this 4-fold degeneracy is actually not universal, which will be explained in Appendix C.

\begin{table*}[!htbp]
\begin{tabular}{|c|c|c|c|}
\hline
Order Parameter & Insulating Phase & Degeneracy & Remaining Anti-Unitary Symmetry \\
\hline
$\sigma^0 \tau^0 \gamma^x$ & VH & 4 & $i\sigma^y \mathcal{K}$, $i\tau^y \mathcal{K}$, $i\sigma^y \tau^x \mathcal{K}$, $i\sigma^x \tau^y \mathcal{K}$\\
\hline
$\sigma^0 \tau^z \gamma^0$ & VP & 2* &\\
\cline{1-3}
$\sigma^0 \tau^z \gamma^x$ & QH & 2 &$i\sigma^y \mathcal{K}$\\
\hline
$\sigma^0 \tau^x \gamma^0$ & IVC & 2 &$i\sigma^y \mathcal{K}$, $i\sigma^y \tau^x \mathcal{K}$\\
\hline
$\sigma^z \tau^0 \gamma^0$ & SP & 2* & \\
\cline{1-3}
$\sigma^z \tau^0 \gamma^x$ & VSH & 2 &$i\tau^y \mathcal{K}$\\
\hline
$\sigma^z \tau^z \gamma^0$ & SVL & 2* &\\
\cline{1-3}
$\sigma^z \tau^z \gamma^x$ & SH & 2 &$i\sigma^y \tau^x \mathcal{K}$, $i\sigma^x \tau^y \mathcal{K}$\\
\hline
$\sigma^z \tau^x \gamma^0$ & SIVCL & 2 &$i\sigma^x \tau^y \mathcal{K}$\\
\hline
\end{tabular}
\centering
\caption{Degeneracy of bound states in different insulating phases. The "*" means bound states emerge only when the strength of impurity potential is small. The remaining anti-unitary symmetries protecting the degeneracy are listed in the last column.}
\end{table*}

\subsection{The Effect of Spin-Orbit Coupling and Zeeman Field}

\begin{table*}[!htbp]
\begin{tabular}{|c|c|c|c|c|}
\hline
Order & Insulating & Degeneracy under & Degeneracy under & Degeneracy under\\
Parameter & Phase & In-Plane Zeeman Field & Perpendicular Zeeman Field & Oblique Zeeman Field\\
\hline
& IVC & 1 & 1 & 1\\
\cline{2-5}
$U(1)_v$ & SIVCL & 2 & 1 & 1\\
\cline{1-5}
& VH & 2 & 2 & 2\\
\cline{2-5}
& QH & 1 & 1 & 1\\
\cline{2-5}
$C_2\mathcal{T}$ & VSH & 2 & 2 & 2\\
\cline{2-5}
& SH & 2 & 1 & 1\\
\cline{1-5}
& VP & 1* & 1* & 1*\\
\cline{2-5}
- & SP & 2* & 2* & 2*\\
\cline{2-5}
& SVL & 2* & 1* & 1*\\
\cline{1-5}
\end{tabular}
\centering
\caption{Degeneracy of bound states in different insulating phases under weak Zeeman fields without SOC. The "*" means bound states emerge only when the strength of impurity potential is small.}
\end{table*}

\begin{table*}[!htbp]
\begin{tabular}{|c|c|c|c|c|}
\hline
Symmetry & Insulating & Degeneracy under & Degeneracy under & Degeneracy under Perpendicular\\
Broken & Phase & No Zeeman Field & In-Plane Zeeman Field & or Oblique Zeeman Field\\
\hline
& IVC & 2 & 1 & 1\\
\cline{2-5}
$U(1)_v$ & SIVCL & 2 & 2 & 1\\
\hline
& VH & 2 & 2 & 1\\
\cline{2-5}
& QH & 1 & 1 & 1\\
\cline{2-5}
$C_2\mathcal{T}$ & VSH & 1 & 1 & 1\\
\cline{2-5}
& SH & 2 & 2 & 1\\
\hline
& VP & 1* & 1* & 1*\\
\cline{2-5}
- & SP & 1* & 1* & 1*\\
\cline{2-5}
& SVL & 2* & 2* & 1*\\
\hline
\end{tabular}
\centering
\caption{Degeneracy of bound states in different insulating phases with weak SOC under weak Zeeman fields. The "*" means bound states emerge only when the strength of impurity potential is small.}
\end{table*}

A spin-orbit coupling (SOC) term reads\cite{Neto:2009aa} $H_{SOC}(\vec{k})=\sigma^z \tau^z \Delta(\vec{k})$, where $\Delta(\vec{k})$ is a two by two matrix in sublattice space. After an SOC term is included, some of the anti-unitary operators in TABLE. I will consequently not commute with the total Hamiltonian, leading to the lift of degeneracy.

A viable way to detect the degeneracy in experiments is to apply a Zeeman field $H_{\text{Zeeman}}=\vec{B}\cdot\vec{\sigma}$ to the TBG and detect the splitting of the degeneracy. Each component of $\vec{B}$ is a two by two matrix representing the strength and direction of the Zeeman field, and $\vec{\sigma}$ is the triplet of Pauli matrices. After a Zeeman field is applied, the degeneracy of the bound states may also be lifted, following the argument above. Without losing generality, Zeeman fields in, perpendicular to and oblique to the plane of TBG are considered. The degeneracy of bound states in different insulating phases subject to Zeeman fields without and with SOC is listed in TABLE II and III, respectively. The dependence of degeneracy on the direction of Zeeman field can be helpful in experiments to the distinction of different insulating phases.

\subsection{The Effect of Magnetic Impurity}

The Hamiltonian of a magnetic impurity in $k$-space has a generic form
\beq
H_{\text{imp}}(\vec{k},\vec{k}')&=&\vec{S}\cdot\vec{\sigma}(\tau^0 \text{Re }U(\vec{k},\vec{k}')+i\tau^z \text{Im }U(\vec{k},\vec{k}')),
\eeq
where $\vec{S}$ is the strength of the magnetic impurity, $\vec{\sigma}$ is the triplet of Pauli matrices and the two by two matrix $U(\vec{k},\vec{k}')$ is the impurity Hamiltonian in sublattice space. In this case, the only candidate of TR-like operators is $i\tau^y\mathcal{K}$. Therefore, only in VH, SP and VSH phase are the bound states 2-fold degenerate. In all of the other phases the bound states are non-degenerate. When SOC is included, there is no TR-like operators commutative to the Hamiltonian, and all the bound states in all the phases are consequently non degenerate.

\section{Conclusions}

In summary, we conclude that the number and property of in-gap bound state peaks is only sensitive to the previously proposed\cite{Lee_2019,pa,op1} symmetry broken by the insulating phases in the TBG. Additionally, we find that the classification of insulating phases via remaining anti-unitary symmetries can further predict the degeneracy of bound states induced by impurities in various conditions and distinguish different phases breaking the same previously proposed\cite{Lee_2019,pa,op1} symmetry. The degeneracy protected by the remaining anti-unitary symmetries depend on only the symmetry of the impurity potential in spin and valley space, but not the detailed form of the impurity potential. By associating the number and properties of in-gap bound state peaks with the classification, all the 9 insulating phases proposed in Ref.\cite{Lee_2019,op1,pa} are distinguishable, as shown in TABLE I, II and III. For an insulating phase whose order parameter is the mixture of two or more order parameters out of the 9 phases, the analysis of remaining anti-unitary symmetries can also dictate the degeneracy of bound states induced by impurities. The in-gap bound state peaks in local DOS and the degeneracy of the bound states can in principle be detected in STM experiments\cite{Kerelsky_2019,Xie_2019,Choi_2019,Jiang_2019,Wong_2020,Choi:2020aa}, and thus can help to the distinction of various insulating phases.

\section{Acknowledgements}

FW and ZQG acknowledge Ashvin Vishwanath for inspiring this work. ZQG thanks Hui Yang and Kai-Wei Sun for enlightening discussions, and thanks Ri-Chen Xiong, Chong-Xiao Fan and Ji-Chen Feng for experimental results helpful to this work. FW acknowledges support from The National Key Research and Development Program of China (Grant No. 2017YFA0302904), and National Natural Science Foundation of China (Grant No. 11888101).

\appendix
\section{Calculation of the Impurity Hamiltonian}

The model proposed in Ref.\citep{b} can be written in the form
\beq
H_{\text{BM}}(\vec{k})=
\begin{pmatrix}
H_{t}(\vec{k}) & T\\
T^\dagger & H_{b}(\vec{k})
\end{pmatrix},
\eeq
where $H_{t(b)}$ is the free Hamiltonian in top (bottom) layer and $T$ is the interlayer hopping. Therefore, a single impurity in top layer reads
\beq
H_{\text{BM,imp}}(\vec{k},\vec{k}')=
\begin{pmatrix}
H_{t,imp}(\vec{k},\vec{k}') & 0\\
0 & 0
\end{pmatrix}.
\eeq
For the impurity potential $U_{\text{imp}}(\vec{r})= u\delta_{\vec{r},\vec{R}_{0}}$, the corresponding impurity Hamiltonian $H_{t,imp}(\vec{k},\vec{k}')$ reads
\beq
H_{t,imp}(\vec{k},\vec{k}')=uC_t^\dagger (\vec{k})C_t(\vec{k}'),
\eeq
where $C_t(\vec{k})=\left(\left| \psi_t(\vec{k})\right>,\cdots \right)$ is constructed by the normalized eigenstates of $H_t(\vec{k})$, i.e. $\left| \psi_t(\vec{k})\right>$. Consider the diagonalization 
\beq
V(\vec{k})H_{\text{BM}}(\vec{k})V^\dagger(\vec{k})=H_0(\vec{k}),
\eeq
where $H_0(\vec{k})$ is given in Eq. \ref{eq:31}. Then the impurity potential projected to the Hilbert space of flat bands reads
\beq
H_{\text{imp}}(\vec{k},\vec{k}')=V(\vec{k})H_{\text{BM,imp}}(\vec{k},\vec{k}')V^\dagger(\vec{k}').
\eeq
In the following calculations involving integration over the MBZ, the mesh of $\vec{k}$ is set to be 1/12 of the length of the MBZ edge.

\section{Numerical Results of the Degeneracy of Bound States}

To examine the degeneracy of bound states, we explicitly calculate the eigenvalues of the total Hamiltonian $H$, whose $(\vec{k},\vec{k}')$-component in $k$-space is $H(\vec{k},\vec{k}')$ given in Eq. \ref{eq:32}. Then we check the degeneracy of the eigenvalues of in-gap states. The numerical results of these eigenvalues are listed in TABLE. IV, which are consistent with the symmetry analysis in Sec. IV.

\begin{table}[!htbp]
\begin{tabular}{|c|c|c|}
\hline
Insulating Phase & Eigenvalues/meV & Degeneracy\\
\hline
VH & -0.00863044 & 4\\
& -0.00863044 &\\
& -0.00863044 &\\
& -0.00863044 &\\
\hline
VP & 0.0579586 & 2\\
& 0.0579586 &\\
\hline
QH & -0.00863044 & 2\\
& -0.00863044 &\\
& -0.00816392 &\\
& -0.00816392 &\\
\hline
IVC & -0.136416 & 2\\
& -0.136416 &\\
& 0.151620 &\\
& 0.151620 &\\
\hline
SP & 0.0579586 & 2\\
& 0.0579586 &\\
\hline
VSH & -0.00863044 & 2\\
& -0.00863044 &\\
& -0.00816392 &\\
& -0.00816392 &\\
\hline
SVL & 0.0579586 & 2\\
& 0.0579586 &\\
\hline
SH & -0.00863044 & 2\\
& -0.00863044 &\\
& -0.00816392 &\\
& -0.00816392 &\\
\hline
SIVCL & -0.136416 & 2\\
& -0.136416 &\\
& 0.151620 &\\
& 0.151620 &\\
\hline
\end{tabular}
\centering
\caption{Numerical results of the degeneracy of bound states. The numerical results are consistent with the symmetry analysis in Sec. IV.}
\end{table}

\section{The Origin of 4-Fold Degeneracy}

First notice that the total Hamiltonian is proportional to $\sigma^0$, which means that the Hamiltonian is the direct sum of two copies of a spin-irrelevant part, namely $\hat{H}=\hat{H}_{\text{spin-ir}}\oplus \hat{H}_{\text{spin-ir}}$. If for $\hat{H}_{\text{spin-ir}}$ there are two linear independent degenerate eigenstates $\left| \psi_1 \right>$ and $\left| \psi_2 \right>$ with the same energy, the four direct summed states $\left| \psi_1 \right>\oplus\mathbf{0}$, $\left| \psi_2 \right>\oplus \mathbf{0}$, $\mathbf{0}\oplus \left| \psi_1 \right>$ and $\mathbf{0}\oplus \left| \psi_2 \right>$ will consequently be four linear independent degenerate eigenstates of the total Hamiltonian $\hat{H}$. Since in each spin-irrelevant Hilbert space the TR-like operator can be chosen as $\mathcal{S}_\tau=i\tau^y \mathcal{K}$, the Kramers theorem ensures that the bound states in each spin-irrelevant Hilbert space must be 2-fold degenerate. Therefore, the total degeneracy for bound states in VH phase is 4.

\bibliography{TBGNote.bib}

\end{document}